# Observational Constraints on the Angular and Spectral Distributions of Photons in Gamma-Ray Burst Sources

V. V. Sokolov[1], G. S. Bisnovatyi-Kogan[2], V.G.Kurt[3], Yu.N.Gnedin[4], and Yu. V. Baryshev[5]

[1]*Special Astrophysical Observatory, Russian Academy of Sciences,
Nizhnii Arkhyz, Karachai-Cherkessian Republic, 357147 Russia*
[2]*Space Research Institute, Profsoyuznaya ul. 84/32, Moscow, 117997 Russia*
[3]*Astro Space Center, Lebedev Physical Institute, Russian Academy of Sciences,
Profsoyuznaya ul. 84/32, Moscow, 117997 Russia*
[4]*Main Astronomical Observatory, Pulkovo, St. Petersburg, Russia*
[5]*Astronomical Institute, St. Petersburg State University, Staryi Petergof, St. Petersburg, Russia*

**Abstract** — The typical spectra of gamma-ray bursts (GRBs) are discussed in the context of the compactness problem for GRB sources and how it is resolved in the popular fireball model. In particular, observational (model-independent) constraints on the collimation of the gamma-rays and the dependence of the collimation angle on the photon energy are considered. The fact that the threshold for the creation of $e^-e^+$ pairs depends on the angle between the momenta of the annihilating photons in the GRB source provides an alternative solution to the compactness problem. A new approach to explaining GRBs, taking into account the angular dependence for pair creation, is proposed, and the main features of a scenario describing a GRB source with a total (photon) energy smaller or of the order of $10^{49}$ erg are laid out. Thus, we are dealing with an alternative to an ultra-relativistic fireball, if it turns out (as follows from observations) that all "long" GRBs are associated with *normal* (not peculiar) core-collapse supernovae. The effects of radiation pressure and the formation of jets as a consequence of even a small amount of anisotropy in the total radiation field in a (compact) GRB source are examined in this alternative model. Possible energy release mechanisms acting in regions smaller or of the order of $10^8$ cm in size (a compact model for a GRB) are discussed. New observational evidence for such compact energy release in the burst source is considered.

## 1. INTRODUCTION

There exists both direct and indirect observational evidence for a connection between massive (core-collapse) supernovae and long gamma-ray bursts (GRBs). The list of publications on this topic is continually becoming more and more extensive. At first, this connection was based on the fact that all the host galaxies of GRBs proved to be forming massive stars at high rates [1–3]. By 2004, more cases were found in which signs of supernovae appeared in the light curves of GRBs or in the spectra of their afterglows (GRB 970228, GRB 970508, GRB 990712, GRB 000911, GRB 021211, GRB 030329, and GRB 031203) [4 - 14]. A thorough analysis of the contribution of supernovae to GRB afterglows was recently undertaken in [15] (see also [16]): in an ever growing number of cases, it is becoming possible to obtain clear photometric and spectroscopic evidence for an association between normal massive supernovae (type Ib/c and other types) and GRBs, which offers direct and incontrovertible proof of a relation between GRBs and massive stars. In turn, the evergrowing statistics on the association between GRBs and supernovae can provide strong observational constraints for the collimation angle of the gamma-rays, and thereby *observational,* model-independent estimates of the intrinsic total energy of the GRB sources [17].

The aim of the current paper is to describe the main assumptions in a scenario for a GRB source with an energy of $\lesssim 10^{49}$ erg. We present both observational and theoretical arguments supporting the idea that classic ("long") GRBs are associated with axially symmetric supernova explosions with the directed ejection of a jet. This approach solves the problem of a fantastically high luminosity of GRB sources ($>10^{51}$ erg), and reduces this energy to reasonable values. Thus, we are considering an alternative to a relativistic fireball model, if we suppose that *all* long GRBs can be associated with *normal* (not peculiar) massive (core-collapse) supernovae.

Section 2 discusses the typical spectra of bursts, the origin of the compactness problem, and how it was solved in the early stages of a GRB. Note that this problem was first formulated in [18] in connection with the giant outburst in the soft gamma-ray repeater SGR 0526-66 on March 5, 1979. Section 3 considers a solution for this problem allowing for the dependence of the threshold for the creation of $e^-e^+$ pairs on the angle between the photon momenta, the degree of collimation or asymmetry (axial symmetry) of the radiation field arising in the source itself, and the dependence of the collimation angle on the photon energy. In Section 4, we discuss the observational basis for strong collimation of the radiation reaching near-Earth receivers from distant ($z \gtrsim 1$) GRBs (but only for a small number of hard photons in the GRB spectra). Section 5 is devoted to jets, which may be formed by the powerful pressure due to the collimated (anisotropic) radiation near ($\lesssim 10^9$ cm) the GRB source. Section 6 considers some possible energy-release mechanisms that could act in a compact GRB source, while Section 7 presents observational evidence for a "compact" ($\lesssim 10^8$ cm in size) region of energy release. In particular, we present a brief analysis of observations (carried out first on BeppoSAX, then on HETE-2 [19, 20]) of soft X-ray flashes (XRFs), GRBs with strong X-ray excesses (X-ray Rich GRBs, XRR GRBs), and normal or classic GRBs. We consider some new observational evidence (superluminal radio components) for compact GRB sources with jets. Finally, we will attempt to understand the soft (in the sense of the photon energy) observed spectrum of GRBs, without assuming *a priori* that the radiating plasma moves with Lorentz factors $\Gamma \gg 10$, as it does in the popular ultrarelativistic fireball model [21, 22].

## 2. TYPICAL GRB SPECTRA AND PHOTON ENERGIES

The rapid time variability of GRBs ($\delta T \sim 10$ ms) implies a *compact* source of radiation, with a size less than $c \times \delta T \approx 3000$ km. However, this immediately raises the problem of the huge luminosities of distant GRB sources (see, for example, [18, 23]): too much energy ($>10^{51}$ erg, even if we consider only the soft gamma-rays, <511 keV and 1 MeV) is released for such a small volume for sources at cosmological distances (>1 Gpc), i.e., for classic, long GRBs. At a photon density of $n_\gamma \sim (10^{51} \text{ erg}/m_ec^2)/(c\delta T)^3 \sim 10^{57}/(3000 \text{ km})^3 \sim 10^{32}$ cm$^{-3}$, *two* gamma-rays with a *total* energy of more than $2m_ec^2$ can interact with each other and give birth to an electron–positron pair. The optical depth to pair creation is given approximately by $\tau_{e-e+} \sim n_\gamma r_e^3 (c\delta T) \sim 10^{16}$, where $r_e$ is the classic radius of the electron, $e^2/m_ec^2$. (For semi-relativistic energies <511 keV up to 1 MeV, the pair-creation cross section is $\sim r_e^2$, i.e., $\sim 10^{-25}$ cm$^2$.) This is the essence of the so-called "compactness problem:" the optical depth for photons with relatively low energies ~511 keV would be so huge that it would be impossible to observe such photons, since they would all be transformed into $e^-e^+$ pairs. Thus, in the presence of such huge photon densities, $n_\gamma$, photons radiated in the GRB source that initially propagate along the line of sight toward the observer should be destroyed by collisions with other photons whose total energy together, with the energy of the primary photon, exceeds $2m_ec^2$ (see Section 3 for the situation when the radiation field in the source is completely isotropic).

However, the usual formulation of the compactness problem [21, 22, 24, 25] immediately reduces to the problem of the escape (from the GRB source) of hard, *high-energy* photons: gamma-ray photons with energies exceeding $2m_ec^2$ (and $\gg 1$ MeV) can interact with a huge number of photons with lower energies (<511 keV to 1 MeV) and create $e-e+$ pairs. The mean optical depth for this process is [22] $\tau_{e-e+} \sim 10^{15}(E/10^{51} \text{ erg})(\delta T/10 \text{ ms})^2$ for a typical total (isotropic) energy release of $E \sim 10^{51}$ erg in a small volume. It is thought that "heavy" (hard, or high-energy) photons are always present in the observed spectra of GRBs, in the form of high-energy "tails" containing an appreciable amount of energy. For example, according to Piran [21, 22, 24, 25], the compactness problem arises precisely because the observed spectrum (probably) always contains some number of hard gamma-rays. In other words, according to Piran, since the observations do not contradict the *possibility* that all GRB spectra have hard tails, this should be the first and main observational basis for the problem in this type of formulation (see, for example, [26]): the optical depth to high-energy, hard photons ($\gg 1$ MeV) should be so great that it will be impossible to observe such photons. However, are such photons really present in all GRB spectra? And does the whole problem have to do precisely with these high-energy photons? Here, we should immediately make some clarifying comments about the typical spectra and typical photon energies of GRBs, based on well-known observational results for GRB spectra (including the most recent ones).

Such photons are indeed observed in *some* cases— but far from all the time. Moreover, photons with energies >100 MeV are strongly delayed relative to the main GRB. For example, the 20-GeV photons (observed on BATSE/EGRET) are delayed by a whole 1.5 h relative to the GRB itself. It is obvious in this case that the physical mechanism operating is completely different from the one that gives rise to typical GRB spectra. (A detailed description of GRB spectra is given in the review [27]; see also the catalog of spectra [28].) A "typical" observed GRB spectra are quite varied, but nevertheless, are composed primarily of soft (and not hard) gamma-rays. This has been known since the very discovery of GRBs, when their spectra were presented in energy units (see, for example, the survey [28]). Now, many authors are pointing out the same thing [30– 35]. Nearly all GRBs were discovered in the energy range from 20 keV to 1 MeV. In his recent survey, Piran [25] himself was forced to note the mystery of the narrow energy range (with its maximum at $E_p <$ 511 keV, [29]) of typically observed GRBs. In addition, it became clear by 2000 that there exist two more classes of GRBs: X-ray Flashes (XRFs) and X-ray Rich GRBs (XRR GRBs) [19, 36]. These are "gamma-ray bursts" involving few or even *no* gamma-rays. This phenomenon is discussed in detail (accompanied by excellent illustrations) by Lamb et al. [30, 31], as well as in other papers published by this same group.

Thus, in spite of the significance of the problem of the emergence of hard (>> 1 MeV) photons, at the same time (and most importantly!), the small volume with radius $R \leq 3000$ km turns out to have too many photons with low energies, but with densities $n_\gamma \sim 10^{32}$ cm$^{-3}$. The observed fluxes give estimates of the total isotropic energy releases ($\gtrsim 10^{51}$ erg) of GRBs precisely in the form of such low-energy photons. In other words, this ("standard," $\sim 10^{51}$ erg) energy release was obtained using typical GRB spectra precisely for these most often obesrved low-energy photons with *semi-relativistic energies,* primarily to 1 MeV. Moreover, it is important to emphasize that both photon density $n_\gamma$ and the total *photon* energy release ($\sim 10^{51}$ erg) were estimated using the simple assumption of spherical symmetry or total isotropy of the radiation field arising in the source during the burst (see below the comments on [23]).

Further, it has been firmly stated [21, 22, 24, 25], many times repeated, and by different authors, that GRB sources could be optically thin, and the observed GRB spectra are non-thermal for sure. Subsequently, an optically thin source with a nonthermal spectrum often appeared as a "standard" general picture for all GRBs [37]. At the same time, in contrast to the GRB spectra that are usually presented in catalogs [27, 28] that are *averaged over the burst time*, *time-resolved* or instantaneous GRB spectra for sufficiently bright bursts are closer to thermal, than power-law, and correspond to blackbody (Planck) radiation with a temperature of $kT \sim$ 100 keV [38–43]. Various authors have pointed out this lack of consistency between the standard optically thin synchrotron model and the observations (see, for example, [40]), and have proposed various alternative scenarios to solve this problem [32, 44, 45]. It is likely that this blackbody radiation with $kT \sim$ 100 keV corresponds to some physical model for the source, while the non-thermal GRB spectra that have been obtained represent only an empirical fit to the *time-averaged* data for observed bursts (see [43] and references therein).

Nevertheless, if these two (more theoretical than observational) assertions, or postulates [21, 22, 24, 25], that (1) it is possible that all GRB spectra have high-energy components and (2) the observed GRB spectra are non-thermal are indeed true, the only possible theoretical alternative is the fireball model, with huge Lorentz factors $\Gamma \gg 10$ [21, 22]. The synchrotron model with a shock and optically thin plasma that is based on these postulates can more or less correctly explain much about GRBs, but not the *observed* spectra of GRBs [40].Somehow, we lose sight of the fact that the "photon targets" [37] in this model (with a typical energy $E_p <$ 511 keV) are the same typical observed GRBs. Thus, it turns out that the main task in the standard fireball model is not to explain the observed spectrum (soft, in the sense of the energy–frequency of the photons), but instead to investigate the rare cases when hard photons with $E \gtrsim 1$ GeV can emerge. In this connection, we must note study [26], in which as an alternative to the observed GRB spectrum, it is proposed to explain some unobserved "true (internal)" GRB spectrum, which doesn't have any cutoff at high energy at all. The origin of the observed, predominantly soft GRB spectra with a large number of photons with energies up to $\sim 1$ MeV remains unclear.

The spectrum is especially unclear against the background of the conjuring about huge Lorentz factors, invoked to solve the compactness problem. However, the question remains: why are primarily soft spectra observed in the presence of the ultrarelativistic motion of the radiating plasma that is assumed in the fireball model? And we shouldn't forget that GRB spectra sometimes do not contain any gamma-rays, as in the case of XRFs, whose existence was known before 2000 [36]. Thus, in solving the compactness problem, we have created another problem of scandalous inconsistency between ultrarelativistic Lorentz factors $\Gamma \sim 100-1000$ (with problematic 100-MeV and 10-GeV photons) and the observed, usually soft ($\lesssim 1$ MeV) gamma-ray (GRB, XRR GRB) and X-ray (XRF) emission of the most "classic" GRBs. Moreover, note that the observed *blackbody* radiation of GRBs with temperatures of $kT \sim 100$ keV [42, 43] is inconsistent with Lorentz factors of $\approx 10^2-10^4$ for the very reason given by Piran: the observed temperature $kT$ in the cosmic fireball [21, 22] could then easily exceed 1MeV [46].

### 3. THRESHOLD FOR THE CREATION OF $e-e+$ PAIRS AND THE ANGLE BETWEEN THE PHOTON MOMENTA

Do there exist other ways to solve the compactness problem, apart from the fireball model with huge Lorentz factors? In particular, can we get away with a semi- (and not ultra-) relativistic approximation when explaining the observed spectra of GRBs, XRR, and XRFs? Is strong "beaming" (collimation) of the gamma-ray radiation needed, and how collimated can the radiation forming the spectra of these types of bursts (GRB, XRR GRB, XRF) be? In this section, we will consider another attempt made to solve an old problem.

Of course, the possibility that the radiation of GRBs was collimated was already being discussed in 1998, but mainly in terms of the "standard" fireball model. Recall that the term "collimation" in this theory refers not to the directly observed gamma-ray emission, but to hypothetical jets of plasma. The term "beaming" in the fireball model refers to the radiation of the optically thin jet plasma [47], which is concentrated in a narrow cone with opening angle $\sim 1/\Gamma$. Below, we will also use the term collimation, but only for the observed radiation; we will not associate it with jets, all the more so with factor $\Gamma$ or with beaming. (The detector registers the gamma-ray burst, not the jet.) In addition to the collimated radiation, we will also consider anisotropic (axially symmetric) radiation, or the radiation field in the immediate vicinity of the GRB source.

There would not be the need to spend so much time discussing the approach of Piran (see Section 2), except for the fact that the compactness problem was actively discussed even before 1991 (i.e., in the pre- BATSE/EGRET era), in connection with the famous burst of March 5, 1979 in the Large Magellanic Cloud [18]. Even then, the possibility of anisotropic or *collimated* gamma-ray radiation was not excluded as an explanation for the soft spectra, since the cross section for the creation (and annihilation) of electron–positron pairs $\sigma_{e-e+}$ depends not only on energy, but also on the angle between the momenta of the colliding particles [18]. We comment below on study [23], which is now cited rather rarely, although it demonstrates that much had been said even in the early 1990s about the compactness problem and the collimation of the radiation emerging from a source with a high photon density. The presence of such anisotropic (collimated) radiation in the source can solve this problem, but in a completely different way (see below). In particular, Carrigan and Katz [23] model the observed spectra of GRBs taking into account $e-e+$ pair creation. This can give rise to efficient collimation of the flux, due to the kinematics of two-photon pair creation: $\tau_{e-e+}$ is sensitive to both the *spectral* and the *angular* distributions of the radiation field in the GRB source.

Due to their importance for our subsequent analysis of the influence of the angular and spectral distributions of the photons on the opacity $\tau_{e-e+}$, we analyze below the formulas for the pair-creation threshold from [23] [see (1)]. A pair can be created by *two* photons with energies $E1$ and $E2$ whose sum exceeds the threshold energy for the creation of $e-e+$ pairs, $2E_{th} < E_1 + E_2$, if

$$E_1 E_2 \geq 2(m_e c^2)^2/(1 - \cos\theta_{12}), \qquad (1)$$

where $2(m_e c^2)^2 = 2(511 \text{ keV})^2$, $\theta_{12}$ is the angle between the directions of the two gamma-rays, and $E_{th} = \sqrt{E_1 E_2}$. As a result of the creation of the pair, one of the photons initially moving along the line of sight toward the observer disappears. For example, a pair will be created during a collision of two photons with oppositely directed wave vectors (i.e., a head-on collision) if the sum of their energies is $E_1 + E_2 > 2E_{th} = 2 \times 511$ keV for $\theta_{12} = 180°$, or, in the case of a side-on collision, if $E_1 + E_2 > 2E_{th} \approx 2 \times 700$ keV for $\theta_{12} \approx 90°$. However, in the case of nearly parallel wave vectors, $\theta_{12} \approx 0°$, the threshold energy for $e-e+$-pair creation $2E_{th}$ becomes very high: in this case, $E_1 + E_2 > 2E_{th}$ when $E_{th} \to \infty$. This means that $e-e+$ pairs will not be created, and the photons can freely leave the source, even when they have energies $\gg 1$ MeV. Further, as was noted in [23], condition (1) leads to softening of the emergent (observed) radiation from the source. If the spectrum of the source does not have a sharp peak, comparatively hard photons ($E > E_{th}$) will create pairs, primarily via their interaction with low-energy photons ($E < E_{th}$). This means that the observed spectrum of a GRB source will be soft, since photons with high energies are prevented from emerging due to condition (1). The $e-e+$ pairs will eventually annihilate, with the creation of two, or rarely three, photons, but usually not in such a way as to give one of the photons high and the other low energy.

Since it is supposed that any *reasonable* source spectrum will contain many more photons with low or moderate energy ($\lesssim 511$ keV) than with very high energy [23], the emergent spectrum will differ more markedly from the initial source spectrum at higher photon energies ($E \gtrsim 1$ MeV), so that the observed spectrum is strongly suppressed at such energies. The high-energy photons are "beaten down" by the large number of low-energy photons. In other words, most photons with high energy are taken out of the line of sight, and therefore out of the observed spectrum, and we should expect to measure a flux of photons with $E > E_{th} = \sqrt{E_1 E_2}$ only if $\tau_{e-e+}$ is less than unity, since the $e-e+$ pair-creation threshold depends on the angular distribution of the radiation field arising in the source.

Thus, there should not be a large number of photons with $E > 1$ MeV and with $E \gg 1$ MeV in the observed spectra of GRBs, due to the influence of (1), as long as the optical depth to the creation of $e-e+$ pairs does not become $\lesssim 1$ (for example, anisotropy of the source radiation field). As we can see from [23], it was usual in 1992 to take the "typical" energies of most photons in the observed spectra of GRBs to be fairly low. Further, Carrigan and Katz [23] estimate the distances to the burst sources in the case of such (observed) semi-relativistic photon energies ($E \sim m_e c^2$). The point is that the compactness problem for GRB sources did arise (in connection with the event of March 5, 1979 in the LMC), given the unexpectedly large distances to these sources—but not because of the "problem" of the emergence of "heavy" (100 MeV, 1 GeV, or more) photons with ultrarelativistic energies ($E \gg m_e c^2$), which are "hindered" by the "light" ($\lesssim 1$ MeV) photons that are observed in GRB spectra. The powerful burst of March 5, 1979 did not have any "superheavy" photons in its spectrum. To convince yourself of this, simply look at the spectrum of this burst published in review [29].

Carrigan and Katz [23] discuss various possible explanations for why the "$e-e+$ confinement" of photons does not operate in the GRB source associated with the March 5, 1979 event in the LMC. In particular, they immediately underscore the importance of taking into account angular dependence (1) for $e-e+$ pair creation. For example, there will be a certain "loop-hole" for the photons if the source itself forms a *collimated* beam of photons. Essentially, we are dealing here with asymmetry or *anisotropy* of the radiation field arising in the source itself during the outburst, as is discussed by Agaronyan and Ozernoi [18]. In this case, even photons with high energies will be below the pair-creation threshold if the angle $\theta_{12}$ is sufficiently small. The presence of such an opacity "window" for collimated photons assumes that even in the region, opaque due to pair creation, the GRB source radiation can escape through this window, by analogy with the large contribution of such windows in the opacity of matter to fluxes of radiation in the usual (Rosseland mean) approximation.

The use of the words "strongly collimated" in [23] might make us pause. Indeed, what is meant by "strongly"? At that time, there were not yet any observations of GRB spectra at high energies $E$. Photons with $E \sim 10$ MeV (beyond the $\sim 1$ MeV peak) began to be reliably observed only with BATSE/EGRET. In particular, we can obtain for such photons from (1) the relation $1 - \cos \theta_{12} = 0.522245$ MeV$^2$/(10 MeV $\times$ 10 MeV) $\approx 0.005$, which corresponds to angle $\theta_{12} < 6°$. Thus, photons with energies $\sim 10$ MeV that leave the source within a cone of opening angle $\sim 6°$ will not create pairs, while the *softer* source radiation can be completely uncollimated. For example, collisions of 10-MeV photons with lower-energy photons ($<100$ keV) occur at angles larger than $60°$ (0.522245 MeV$^2$/(10 MeV $\times$ 100 keV) $\approx 0.5$), while the softer photons emerging from the source within a cone with this opening angle will not prevent either heavy or, *all the more so,* light photons from freely leaving the source. Thus, Eq. (1) requires more or less strong collimation only for a *small fraction* of the hardest photons emitted by the source. This becomes clear if we consider the energy spectra of typical GRBs [29] presented in the old form of $F$ (cm$^{-2}$ s$^{-1}$ keV$^{-1}$) plotted against $E$ (keV), i.e., in terms of the *number* of photons per unit time per unit energy interval per unit area as a function of the energy. Only a small fraction or small number of the photons that are observed beyond the threshold at ≈700 keV can be "strongly" collimated, within a cone with opening angle <90°.

The angles ~6° are now considered quite appropriate for the jet opening angles in the standard, or, more precisely, most popular, the fireball model. If we start from the idea that we want to enable the emergence of photons with energies to 10 keV from the source, we can immediately obtain a type of "collimation theory" with $\Gamma \sim 10$. However, this path is also dead end within the standard fireball model. Allowing for initial collimation of the radiation arising in the source, rather than invoking collimation due to huge Lorentz factors ($\Gamma \sim 1000$ is required to solve the compactness problem), can completely change the entire model.

On the other hand, one way or another, the fluxes of photons from the source should lead to effects due to the pressure of the radiation on the material surrounding the source. If the radiation is also collimated, the formation of jets becomes unavoidable in the presence of such *huge* fluxes, if there is even a small amount of asymmetry (or anisotropy) of the GRB-source radiation field. This is precisely the question: is the source of the GRB a jet?

## 4. OBSERVATIONAL ESTIMATE OF THE COLLIMATION ANGLE FOR GAMMA-RAYS REACHING NEAR-EARTH INSTRUMENTS

In fact, perhaps angular dependence (1) for the $e+e-$ pair-creation threshold should have been taken into account right from the start, without assuming ultrarelativistic motion of the plasma early on, and allowing for the possibility of special directed in the source of a flare at the surface of a compact object— the GRB source?—due to the action of some agent (most likely magnetic fields). A "special direction in the source" sounds as a challenge. However, the emitting plasma in a fireball model with jets must also somehow be accelerated to huge kinetic energies, but by what mechanism? This question remains unanswered in the theory of ultrarelativistic fireballs, just as the origin of the observed spectra of GRBs remains unclear in this theory (see Section 2). Does the (ultrarelativistic) jet itself radiate, and is it actually the source of the GRB? This is the question—and can we manage without this gamma-ray radiating jet that must be accelerated (by some unknown means) to huge Lorentz factors, by supposing that the radiation of the burst source is *already* collimated by the source itself?

In any case, the idea that the gamma-ray radiation reaching near-Earth detectors might be strongly collimated can be given an observational basis (without worrying for the moment about the collimation mechanism), if we suppose that all long GRBs are associated with the explosions of *ordinary* massive (core-collapse) supernovae [3, 48]. Here, again, we emphasize that we are imagining collimated radiation, and not a jet. It is precisely the radiation that emerges from the source that is then detected. The formation of jets (moving with speeds less than the speed of light) will be considered in the following section.

So far, all available photometric and spectral observations of the host galaxies support the existence of a relationship between GRBs and the evolution of massive stars, or a close connection with the relativistic collapses and supernova explosions that occur at the end of their evolution. Many papers have dealt with this topic, such as [1–3] and others. The main conclusion reached by these studies is that the host galaxies of GRBs are basically indistinguishable from other galaxies with similar redshifts $z$, in terms of their colors, spectra, star-formation rates, luminosities, and surface brightnesses. Thus, they are likely star-forming galaxies that are quite ordinary for their $z$, and comprise the main basis for samples made from deep surveys of weak objects (at least surveys in the optical to $26^m$, where bursts of star formation are visible to $z \gtrsim 1$).

In essence, this is the main result of the optical identification of GRBs with objects of a known nature: long bursts are identified with ordinary, *nonpeculiar* galaxies of types that are frequently encountered at their $z$, at least down to ≈$26^m$. This means that we can use counts of such galaxies (the number of galaxies brighter than $26^m$) and the results of direct optical identifications of GRBs (detected to $\gtrsim 10^{-7}$ erg/cm$^2$) to estimate the mean yearly rate of GRBs occurring in each such galaxy. This rate turns out to be $N_{GRB} \sim 10^{-8}$ yr$^{-1}$, on average, in each star-forming galaxy. Since the yearly rate of (massive, core-collapse) supernovae is $N_{SN} \sim 10^{-3}–10^{-2}$ yr$^{-1}$, the ratio of the GRB rate to the rate of such supernovae is close to $N_{GRB}/N_{SN} \sim 10^{-5}–10^{-6}$. This estimate (most likely an upper limit) was made for type-Ib/c supernovae [48]; the analogous estimate for type-II supernovae was obtained by Porciani and Madau [49]: $(1–2) \times 10^{-6}$.

Further, we will start with a very simple assumption, that has been supported by a larger number of observational facts since 1998 [15, 16]: *all* long GRBs are associated with the explosions of massive supernovae. The ratio $N_{GRB}/N_{SN}$ should then be interpreted as reflecting very strong collimation of the gamma-rays *reaching the observer;* i.e., the gamma-rays from the source (or some fraction of them) propagate to large distances ($z \gtrsim 1$) within a very narrow solid angle

$$\Omega_{beam} = N_{GRB}/N_{SN} \times 4\pi \sim (10^{-5}–10^{-6}) \times 4\pi. \quad (2)$$

Another possible interpretation of the small value of $N_{GRB}/N_{SN}$ is that the GRBs are associated with some rare class of peculiar supernova (hypernovae?). We consider this to be less likely, because GRBs are then associated with only $10^{-5}–10^{-6}$ of all supernovae observed thus far. The counterparts of GRBs are not simply peculiar supernovae with which the "Pachinsky hypernova" is sometimes identified [50, 51], since these peculiar supernovae (hypernovae), such as 1997ef, 1998bw, and 2002ap, are too few in number (see [52, 53]) for such a small ratio $N_{GRB}/N_{SN}$. There are also other studies [20, 30, 31] suggesting the possible collimation of the GRB source radiation with the collimation angle (2). The more GRB–SN coincidences such as that between GRB 030329 and SN 2003dh or the correlation between GRBs and the "red shoulder" in the light curves [15, 16], the more certain it will become that the GRB radiation is indeed collimated, rather than being associated with some special class of supernovae, such as hypernovae. Many believe that the very term "hypernova" is poorly defined and have ceased to use it (see, for example, [50]). The geometry of "ordinary" supernova explosions (which can be axially symmetric) further complicates attempts to distinguish some special class of "hypernovae" [54]. It is likely that, in the case of GRB–SN coincidences, we are observing an axially symmetric supernova explosion very close to some special direction— the axis of the explosion, which leads to peculiar photometric and spectral properties in ordinary type- Ib/c supernovae (high luminosities, high velocities, etc.).

Let us suppose that a distant observer (for $z \gtrsim 1$) receives only the most collimated part of the gamma-ray radiation, for example, because his line of sight lies along the rotational axis of the collapsing core of a star with a magnetic field. If the gamma-rays are strongly enough collimated that they illuminate only a small fraction of the sky, the inferred energetics of the event as a whole must be strongly decreased to several orders of magnitude below the so-called "isotropic equivalent" total energy release of the GRB source $E_{iso}$. This quantity usually has values $E_{iso} \sim 10^{51}–10^{52}$ erg, but can reach $\sim 10^{53}$ erg. The true total GRB energy will be $\sim E_{beam}$, where

$$E_{beam} = E_{iso}\Omega_{beam}/4\pi, \quad (3)$$

or $\sim 10^{45}–10^{47}$ erg.

If this situation is realized for some fraction of the gamma-rays emitted by the source (carrying from $\sim 10^{47}$ to $\sim 10^{49}$ erg) that propagate within a narrow beam (2) that reaches an observer on Earth some other part of the source energy could be radiated *isotropically* (or nearly so). However, if the spherically symmetric luminosity corresponds to a total isotropic GRB energy of, for example $\sim 10^{45}–10^{47}$ erg, even a detector as sensitive as the BATSE GRB-monitoring instrument would not be able to detect the radiation flux from such a low gamma-ray luminosity emitted at cosmological distances, $z \gtrsim 1$, so that the GRB would not be detected if the observer were outside the cone (2) for the collimated component of the radiation. Thus, an energy of $10^{45}–10^{47}$ erg could be close to a *lower limit* for the total (radiated) energy of GRB sources corresponding to the fluxes measured within solid angles $\sim \Omega_{beam}$, within which the most collimated components of the GRB radiation propagate. (Of course, this is valid only if *all* long GRBs are associated with supernovae.)

Thus, based on known observational results, there exists this possibility to immediately appreciably lower the total (bolometric) energies of GRBs, even in models with radiating plasma, i.e., with relativistic jets, as in the standard fireball model. (Although we can be certain that we are not dealing with radiating rapidly moving plasma, therefore the jet is not the origin of the GRB. See below for more detail.) In a model with a radiating jet, we can estimate the Lorentz factor using the formulas relating the energy with $\Gamma$ [21, 22]. For the total energy releases of $10^{45}$, $10^{47}$, $10^{48}$, and $10^{49}$ erg, we obtain for Γ the values 18, 32, 42, and 56, respectively. Such Γ values could correspond to estimates of the collimation angle for photons with energies ~10 MeV (see the end of Section 3), but such low burst energies are completely unacceptable in the "standard" solution of the compactness problem [21, 22, 24, 25].

This is indeed a fundamental question for the GRB mechanism: what is actually radiating—a "point" or an extended jet? If the GRB radiation (mainly the hard component) is collimated, we must return to the idea that the GRB radiation arises at the *surface* (more precisely, near the surface—at distances of centimeters, or meters?) of some compact object. Further, we will proceed without the *a priori* assumption that only the "end" of the jet itself should radiate. First some radiation or the GRB arises, then the jet, as a consequence rather than the origin of the GRB. A jet probably does form, but as a result of the strong pressure of the collimated radiation on the matter surrounding the compact GRB source (with a size of $10^7$ cm or less). Of course, this jet, which has been accelerated to relativistic speeds by the incident photons, will also radiate, but this represents the GRB afterglow, and not the GRB itself.

## 5. RADIATION PRESSURE AND THE JET IN A COMPACT GRB MODEL

If a scenario such as *massive star* → Wolf Rayet (WR) *star* → *pre-SN* = *pre-GRB* → *collapse of massive stellar core* is possible, with a dense envelope being present around the WR star, a relativistic jet could be formed by the powerful pressure of the collimated or anisotropic radiation of the GRB source (i.e., the prompt GRB radiation) on the material in the WR envelope surrounding the source—the collapsing core of the WR star.

We can put aside the question of the mechanism for the formation of the GRB source itself, and not worry about how the collimated gamma-ray beam arose. The radiation field generated around the source could be anisotropic, for example, axially symmetric, due to magnetic fields, inhomogeneity on the surface of the source (e.g., polar caps), or the effect of the angular dependence (1) of the $e-e+$ pair-creation threshold. It is enough to allow for the possibility that either some fraction (~10%, or even 1%) or all the energy of the GRB (~$10^{47}$–$10^{49}$ erg) could be in the form of collimated radiation "tearing" through the dense envelope surrounding the collapsing core of the WR star. This (prompt) radiation then reaches the Earth and is detected as a GRB. The main necessary elements here are the presence of a *directed flow* of radiation from the source and the possible existence of a *dense gaseous* (wind) medium, onto which the radiation of the GRB source immersed in this medium impinges. This medium could plausibly be densest near the source, if the distance dependence of the density displays the typical "WR law" for stellar wind: $n = Ar^{-2}$ (where the distance $r$ is measured from the center of the WR star and $A \sim 10^{34}$ cm$^{-1}$ [55]).

The light-pressure force acting on the gaseous medium (plasma) surrounding the GRB source (WR core) will be $L_{GRB}(4\pi r^2)^{-1}(\sigma_T/c)$, where $L_{GRB}$ is the so-called isotropic-equivalent *luminosity* of the GRB source (usually ~$10^{50}$–$10^{51}$ erg/s or more), $r$ the distance from the center (source), $\sigma_T = 0.66 \times 10^{-24}$ cm$^2$ the Thomson cross section, and $c$ the speed of light. Even without detailed calculations, it is clear that, near the WR core ($r \sim 10^9$ cm), this force can exceed the light-pressure force corresponding to the Eddington luminosity (~$10^{38}$ erg/s for $1M_\odot$) by many (up to 12–13) orders of magnitude.

In principle, isotropic radiation with huge luminosity $L_{GRB} \sim 10^{50}$–$10^{51}$ erg/s could lead to the rapid acceleration (or detonation) of the medium adjacent to the source. However, if we suppose that the GRB-source radiation is anisotropic, with some fraction propagating within solid angle $\Omega_{beam} \sim (10^{-5}$–$10^{-6}) \times 4\pi$, the formation of a relativistic or ultra-relativistic jets becomes inevitable due to the gigantic light pressure acting on the *dense gaseous* medium surrounding the source. Of course, details of the generation of the jets will depend on the degree of ionization, density, and temperature of the plasma in the immediate vicinity of the axially symmetric collapsing core of the massive star that is the GRB source [56]. Here, we will only present estimates of the sizes of regions within which such a jet could be accelerated to relativistic speeds by the incident light pressure.

1. If the flux of photons that is exerting the light pressure accelerating matter at a distance $r$ from the GRB-source center (somewhere near the GRB source, $r \stackrel{>}{\sim} 10^9$ cm) is equal to $L_{GRB}(4\pi r^2)^{-1}$, this flux could be enormous in the immediate vicinity of the collapsing WR core. It is precisely *within* this region that jets are formed and accelerated to relativistic speeds ≈$c$.

2. At the *outer* boundary of this region, the flux of photons must be no lower than the Eddington

flux, $L_{Edd}(4\pi R_*^2)^{-1}$, in order to accelerate the matter to speeds of at least ~0.3 $c$. Here, $L_{Edd}$ is the Eddington luminosity (~$10^{38}$ erg/s for 1 $M_\odot$) and $R_* \sim 10^6$ cm is the size of the compact object. $L_{Edd}(4\pi R_*^2)^{-1}$ is the radiation flux that would cause accretion onto the compact object to cease; i.e., for matter to fall onto the object with the parabolic velocity. For a neutron star, this velocity is ≈0.3 $c$.

Using the condition that the photon flux $L_{GRB}(4\pi r^2)^{-1}$ be equal to (or at least not less than) $L_{Edd}(4\pi R_*^2)^{-1}$ at some distance $r$ and the fact that the GRB luminosity (more precisely, its isotropic equivalent) is $L_{GRB} \sim 10^{50}$–$10^{51}$ erg/s, we can derive the characteristic size ~$10^{12}$ cm ≈ $14 R_\odot$. At this outer boundary, the light pressure is still able to accelerate an initially stationary medium to appreciable velocities ~0.3$c$. *Still deeper*, at distances less than ~$10^{12}$ cm from the source (say $r \sim 10^9$ cm, somewhere *inside* the region initially occupied by the collapsing core of the massive star), the light pressure accelerates the matter to relativistic velocities (with Lorentz factors ~10) when $L_{GRB} \sim 10^{50}$ erg/s. Thus, the formation of a jet could occur in a fairly small volume with characteristic size $\lesssim R_\odot$, in agreement with the observations of a variable absorption feature that formed simultaneously with the GRB in the BeppoSAX/WFC spectrum of GRB 970508 [57, 58]. Thus, a relativistic jet arises somewhere within a region no larger than 10–15$R_\odot$ in size, as a result of the strong light pressure exerted by the GRB radiation on the dense medium surrounding the source.

Of course, the deceleration of such a jet in the circumstellar medium must be considered separately. However, it is possible that there will not be strong deceleration of the jet due to the interaction of the resulting relativistic shock with the surrounding medium (as in the fireball model), even when the medium surrounding the WR star has a high density (to $n \sim 10^{10}$ cm$^{-3}$ for $r \sim 15R_\odot$), because the appreciably more compact relativistic jet (or "bullet") will be decelerated and not the shock, as should be the case in the fireball model [59–61]. Here, we are not dealing with a broad "bulldozer-like" shock (a jet with opening angle ~10°) that sweeps up material in front of itself, and there is essentially no deceleration of the bullet as it moves in the direction of decreasing density of the medium around the massive core of the WR star (with $n = Ar^{-2}$). Therefore, due to the small amount of deceleration and the small radiative losses together with its high initial momentum, the bullet can move at relativistic speed with a Lorentz factor of $\Gamma \sim 10$ the entire time the GRB afterglow is observed; i.e., during the extent of its entire light curve, with its various maxima (or breaks). Of course, here as well, the shock that arises during the motion of the jet–bullet through the circumstellar envelope should heat this medium and then radiate (in the X-ray, optical, and radio) in places where the medium is sufficiently dense. This will inevitably lead to variability ("humps" and breaks in the light curve) of the GRB afterglow due to the radiation by inhomogeneities in density distribution $n(r)$ at distances from the source of $\lesssim 10^{15}$ cm[17], formed by the stellar wind of the parent star.

Below, we list the main assumptions of the scenario describing a GRB source with a (radiated) energy of $10^{47}$–$10^{49}$ erg in the partially filled space around the massive parent star.

(1) At distances from $\lesssim 10^9$ cm (the size of the core of the massive star) to $r \sim 10^{15}$ cm or more, where the WR wind begins to interact with the circumstellar medium in the gas- and dust-filled star-formation region in which the WR star was born, there is a dense (wind) medium that forms an envelope, which resulted due to the evolution of the massive star.

(2) A huge radiation pressure leads to the formation of a jet in the region from ~$10^9$ cm to ~$2 \times 10^{11}$ cm, where the envelope density is the highest (~$10^{15}$–$10^{10}$ cm$^{-3}$), but the optical depth to Thomson scattering can already be less than unity ($\tau \sim \sigma_T n r < 1$).

(3) The outburst itself (possibly a nearly spherically symmetric "gamma-ray explosion") with a total energy of up to ~$10^{49}$ erg arises somewhere closer to the center of the star, within a region ~$3 \times 10^9$ cm in size, or even smaller ~$10^8$–$10^6$ cm, where the WR law for the stellar-wind density, $n = Ar^{-2}$, is not satisfied. It is possible that the explosion (outburst) occurs directly on the surface of the compact object that arises due to the collapse of the massive core.

(4) Only the most collimated part of the GRB source radiation propagating within a solid angle, $\Omega_{beam} \sim (10^{-5}$–$10^{-6}) \times 4\pi$ sr, reaches infinity. The total radiated source energy is either of the same order of magnitude as $E_{beam} = E_\gamma \Omega_{beam}/4\pi \sim 10^{45}$–$10^{47}$ erg, or is about $10^{49}$ erg.

We can now return to the collimation of the photons arising at the surface of the compact object. The creation of $e-e+$ pairs will not disturb the free escape to infinity of photons whose wave vectors lie within a cone with opening angles $\theta_{12}$ such that (1 − cos $\theta_{12}$) < $2(m_e c^2)^2/(E_1 E_2)$. The pair-creation threshold $E_{th} = \sqrt{E_1 E_2}$ within this cone is very high for small $\theta_{12}$ ($\sqrt{(E_1 E_2/2)} \gg 511$ keV), and all photons with energies below this threshold can freely escape through this window of transparency for the collimated photons. The initial spectrum of the GRB source [before filter (1)] is nearly the same as the observed spectrum. The largest changes in the observed spectrum due to the action of filter (1) occur in the hard part of the spectrum, as is discussed above [23]. It is not necessary to try to "invent" some special mechanism leading to the very high collimation (channelization) of all the photons. For example, the radiation field arising around the source could be anisotropic, e.g., axially symmetric due to the local magnetic field. In particular, non-uniform radiation at the source surface (e.g. polar caps) could lead to efficient collimation or anisotropy of the radiation field, due to the influence of the angular dependence (1) for the $e-e+$ pair-creation threshold [23]. Such anisotropy could be associated with the transport of radiation in a medium with a strong magnetic field, when the absorption coefficient for photons polarized orthogonal to the magnetic field (the extraordinary wave) is very small [63, 64].

There is no doubt that the GRB radiation must be collimated somehow, but collimation (2) is relevant mainly for a small fraction of the hard photons. The pair-creation threshold for such photons naturally and smoothly (in proportion to $1 - \cos\theta$) increases as the angle at which a photon is radiated from the surface of the compact object relative to some *distinguished* direction (e.g., the direction of the surface magnetic field) decreases. As a result, *in addition to the soft component,* more hard radiation in the GRB spectrum is transmitted, giving rise to an anisotropic (axially symmetric) radiation field around the source. In particular, it then becomes clear why XRFs and XRR GRBs are completely uncollimated or, most likely, nearly isotropic, with energies of the order of $\gtrsim 10^{49}$ erg [20].

What characteristic sizes are adopted in the standard model? The earliest stages of a GRB are considered in [62]. The GRB afterglow is due to radiation arising because of deceleration of the shock, which begins at $R_{dec} \sim 10^{15}-10^{17}$ cm. This depends on the density of the surrounding medium and the initial Lorentz factor. Thus, $R_{dec}$ is the size of the fireball before the onset of deceleration, and does not exceed $10^{17}$ cm. In the fireball theory, this is essentially the size of the zone where the prompt radiation of the GRB source with the observed spectrum arises: $\sim 10^{15}-10^{17}$ cm. Further begins the zone of the GRB afterglow, whose onset is minutes, hours, or days after the GRB itself.

As was noted above, allowing for the dependence of the $e-e+$ pair-creation threshold on the angle between the photon momenta in the source makes it possible for the GRB radiation to arise in a region that is nearly ten orders of magnitude smaller. The physical conditions leading to the energy release of the GRB source should then be completely different (than for the ultra-relativistic fireball).

## 6. POSSIBLE ENERGY-RELEASE MECHANISMS IN A COMPACT GRB-SOURCE MODEL

Of course, we must ultimately analyze the GRB scenario proposed here with regard to some specific physical mechanism for the development of a GRB at or very near the surface of an object such as a neutron star or quark (strange) star. However, we must first consider various possible types of *energy release* or explosions associated with such objects. Below, we outline some mechanisms that were proposed earlier, that are capable of providing the required GRB energy in a compact model (see also the reviews [65, 66]).

A mechanism for the generation of a GRB during a supernova explosion due to the action of a pulse of neutrinos was proposed in 1975 [67]. Earlier, Bisnovatyi-Kogan and Chchetkin [68] showed that a layer is made out of chemical equilibrium forms in the crust of a neutron star with $\rho = 10^{10}-10^{12}$ g/cm$^3$, which is stable at such high densities. The outward mixing of nonequilibrium material can come about as a result of a star-quake, leading to rapid nuclear detonation due to the fission of super-heavy nuclei. A shock forms, which heats the surface layers to high temperatures, $\sim 10^9$ K ($kT \sim 100$ keV), and the hard radiation of the GRB can arise due to delayed fission reactions. This scenario was proposed for Galactic GRBs (see also [69]), but can now be applied anew in our compact model for a cosmological GRB.

In the 1980s, a mechanism for a GRB source in the vicinity of a collapsing object based on neutrino–antineutrino annihilation was proposed [70]. The efficiency of transforming the energy of a flux of neutrinos, $W_\nu \sim 6 \times 10^{53}$ erg, into a burst of X-ray and gamma-ray radiation is $\alpha \sim 6 \times 10^{-6}$, and the total energy released in the GRB is $W_{X,\gamma} \sim 3 \times$ $10^{48}$ erg. This corresponds to the total GRB energy of $\sim 10^{47}$–$10^{49}$ erg in the compact model with (2).

The subsequently well-studied magneto-rotational explosion mechanism for energy release in the vicinity of a collapsing object was first proposed in 1971 in connection with supernova explosions [71]. Numerical computations of the collapse and subsequent explosion of a magnetized rotating cloud, or the evolved core of a massive star (pre-supernova), showed that the efficiency for the transformation of rotational energy into the kinetic energy of the explosion can be roughly 10% [72–74]. These results were obtained for two-dimensional computations of a magneto-rotational explosion that occurs due to the development of processes in a strongly magnetized, rapidly differentially rotating star. The energy released is sufficient for a supernova explosion, but probably not for the total X-ray and gamma-ray energy of a GRB in the standard fireball model ($\sim 10^{51}$–$10^{53}$ erg). Thus, the possible existence of a strong global magnetic field in the vicinity of a collapsing object, accordingly in the region of a (compact) GRB source, has been discussed in many studies (see also [75–78]).

In this connection, we note also the possibility of a compact release of energy or appearance of a GRB as a result of the decay of a strongly magnetized vacuum around a neutron star with such a field. The idea that this mechanism might be able to explain GRBs was first expressed by Gnedin and Kiikov [79], who also presented the first estimates of the energy stored in the strongly magnetized vacuum, which is quite sufficient to provide the required energy release via the decay of this vacuum. The source energy in the compact model, $10^{47}$–$10^{49}$ erg, corresponds well with the vacuum energy around a newly born neutron star with a super-strong field of $B \sim 10^{15}$–$10^{16}$ G (provided the stellar surface undergoes oscillations, which can play the role of triggering the mechanism for this decay [80]). The possibility of vacuum decay in a super-strong magnetic field has been actively discussed recently [81–84]. However, the main question remains on how the energy of the magnetized vacuum is transformed into radiation.

Gigantic flares from the so-called soft gamma-ray repeaters SGR 0526–66, SGR 1900+01, and SGR 1806–20 were observed on March 5, 1979, August 27, 1998, and December 27, 2004, respectively. The maximum luminosities during these flares were huge, reaching $\sim 10^{47}$ erg/s [29, 85, 86]. Recall that the compactness problem arose precisely in connection with the high luminosity and very short time scales for the burst rise and variability (to $10^{-3}$ s) of the event of March 5, 1979 [18, 23]. The high flare activity of soft gamma-ray repeaters may be due to rapid heating of the naked surface of a strange (quark) star, with the subsequent radiation of a *thermal* ($kT > 100$ keV) emission [87, 88]. The heating mechanism could be, for example, the rapid decay of super-strong ($\sim 10^{15}$–$10^{16}$ G) magnetic fields. This same mechanism can be used to explain long (cosmological) GRBs in our compact model with a collimated gamma-ray component (2). In this case, a prolonged GRB can be considered a set of short flares, similar to the gigantic flares of March 5, 1979, August 27, 1998, and December 27, 2004, with the total duration of the GRB being up to $\sim 10^2$ s.

It is clear from the above that attempts GRBs, using the physics of massive compact objects, have a long history. This experience can be used to work out in detail a compact model for a GRB, or a scenario with compact energy release, taking into account the observational constraints and theoretical arguments in favor of such a path (with a burst energy of $\sim 10^{47}$–$10^{49}$ erg) for solving the compactness problem for GRB sources.

## 7. CONCLUSION. OBSERVATIONAL CONSEQUENCES OF THE COMPACT GRB MODEL

In a compact GRB model, XRFs could be completely uncollimated or poorly collimated (like XRR GRBs), but have low total (bolometric) energies of $\sim 10^{47}$ erg. Since these are most likely the explosions of massive supernovae at distances of up to 100 Mpc [89, 90], we would expect to observe them much more often than in the fireball model. It is important to try to find early spectral and photometric signs of supernovae in the afterglows of such events. Generally speaking, the observational task of identifying XRFs/XRRs/GRBs then becomes a special subsection of the studies of cosmological supernovae. (Recall that GRB 030329/SN 2003dh is an XRR, and not a classic GRB.)

With regard to normal (classic) GRBs, and especially those (with $z \gtrsim 1$) whose spectra contain many "heavy" photons, we can derive a (kinematic) estimate of the limiting collimation angle for the gamma-ray radiation directly from (1), which (independently!) agrees with the observed ratio of the yearly rates $N_{GRB}/N_{SN} \sim 10^{-5}$–$10^{-6}$. If we consider photons with $E \sim 100$ MeV in the most distant GRBs, it follows from the relation $1 - \cos\theta_{12} \approx 0.5$ MeV$^2$/(100 MeV × 100 MeV) = $0.5 \times 10^{-4}$ that the radiation of such GRBs will be the most collimated. Such photons should be radiated into a cone with an opening angle of ≈0.5°, and be detected in the spectra of only fairly distant GRBs (with $z \sim 1$ and further), purely due to this geometrical factor.

Thus, a natural observational consequence of our compact model for GRB sources is that distant GRBs ($z \gtrsim 1$) should be harder, while nearby GRBs ($z \sim 0.1$) will resemble XRFs or XRR GRBs, whose spectra are dominated by soft X-ray photons (although the spectrum will also be affected by the cosmological factor of $1 + z$). Of course, we must also allow for observational selection effects due to the finite sensitivites of GRB detectors. For example, the soft component of distant (classic) GRBs can be "cut off" by the sensitivity threshold of the detector. In this case, the isotropic X-ray burst emitted simultaneously with the GRB may simply not be detected in distant (classic) bursts due to the low-total (bolometric) luminosity of the source ($<10^{49}$ erg). Indeed, XRFs and XRR GRBs all have lower energies than the isotropic-equivalent luminosities $E_{iso}$ of GRBs [19, 20, 30, 31].

As a result, the only radiation that is important for observations with GRB detectors is the radiation propagating within a small solid angle near some distinguished direction: the "remnants" of the soft radiation for XRR GRBs (which somehow manages to exceed the detection threshold of the instrument used) and hard, or even "heavy," photons radiated by the source up to the pair-creation threshold (1) in classic GRBs with $z \gtrsim 1$. (For a specified detection threshold, there may be no gamma-rays in the spectra of nearby XRF-type bursts with $z \lesssim 0.1$.) Although it follows from review [91] that "typical" GRBs are observed at 30 keV–100 MeV, it turns out (as has long been known) that most bursts are much softer (see Section 2). The special importance of this *observational* result of the BeppoSAX and HETE-2 missions is emphasized in [20, 30, 31]. We are referring here to the detection of XRFs, XRR GRBs, and GRBs with clearly different spectra and luminosities, first by BeppoSAX [19], and then by HETE-2. Thus, our compact model can explain the actively discussed Amati relation between $E_p$ and $E_{iso}$, where $E_p$ is the energy corresponding to the maximum energy release (the typical energy) in the spectra of XRFs, (XRR GRBs), and GRBs. In our compact GRB source model, the Amati relation can be a "simple" consequence of both (1) and the anisotropy of the radiation field, most likely associated with the magnetic field at or near the surface of the compact object. As we indicated above, the anisotropy could be associated with the transport of radiation in a medium with a strong (regular, $\sim 10^{14}$–$10^{16}$ G) magnetic field, so that the absorption coefficient for photons polarized orthogonal to the magnetic field (the extraordinary wave) is very small [63, 64]. In this case, the observation of strong linear polarization of the GRB radiation should be another consequence of our compact GRB model.

In the jet-formation scenario considered here, which was used to interpret the light curves of the optical transient GRB 970508 and the X-ray spectra of this GRB's afterglow [17], the X-ray, optical, and radio afterglow emission can be isotropic. There is evidence that the X-ray emission was, indeed, isotropic [92–95]. The initial assumption was the possibility of the small GRB collimation angle (2), which follows from a comparison of the rates of GRBs and supernovae in distant galaxies. Thus, the basic assumption was that there was a close relationship between GRBs and supernovae. *All* long GRBs accompany supernovae, but sometimes they are observed, and sometimes they are not [15, 16, 48, 58]. In other words, long GRBs are associated with the collapse of a massive star or an initially *axially symmetric* supernova, and GRBs should always accompany type Ib/c or other types of massive supernovae. The total gamma-ray energy released by the GRB source should then, in any case, be *no more than,* the entire radiative energy of a supernova ($\simeq 10^{49}$ erg). (It is interesting that the total X-ray energies for GRB 970508, GRB 970828, GRB 991216, and GRB 000214 observed by BeppoSAX, ASCA, and Chandra are of the same order of magnitude; see the master data in [96].)

With such low total energies for GRBs, the only possibility for detecting GRBs at cosmological distances is detecting some strongly collimated fraction of this energy (1−10%) that leaves the source within a solid angle $\Omega_{beam} \sim (10^{-5}$–$10^{-6}) \times 4\pi$. The remaining radiation may simply not be accessible to a GRB detector with a sensitivity of $\sim 10^{-7}$ erg s$^{-1}$ cm$^{-2}$. Of course, this was not true of the much more sensitive (by a factor of 10 000) X-ray telescopes on board the Ariel V, HEAO-1, and Einstein observatories, which were used to carry out the all-sky surveys [36]. With the limiting sensitivity of $\sim 10^{-11}$ erg s$^{-1}$ cm$^{-2}$ at 0.2–3.5 keV, the Einstein X-ray Observatory detected short, unidentified bursts of X-rays similar to GRBs distributed throughout the sky (Fast X-ray Transients) at a rate of $2 \times 10^6$ yr$^{-1}$. This is in good agreement with the mean rate of massive supernova explosions in distant galaxies, but, due to their limited sensitivities, present-day GRB detectors are able to detect only $\sim 1/10\,000$ of this large number of distant supernovae in the form of (long) GRBs.

Our proposed scenario for a compact GRB also enables us to predict the behavior of superluminal radio components, such as those recently observed for GRB 030329 [97]. As we noted in Section 5, there is likely no appreciable deceleration of the narrow jet (bullet) moving with a Lorentz factor of the order of ten. If this is the case, we expect superluminal radio components associated with the jet to have the following properties:

(1) such components observed in the afterglow of a GRB should move with a constant speed;

(2) the characteristic observed speed should be of the order of the Lorentz factor for the jet motion, $\sim 10\,c$.

We will make two final comments.

1. It stands to reason that, with a total (bolometric) burst energy of $\sim 10^{47}$–$10^{49}$ erg and radiation in the narrow cone (2), the GRB energy (3) for the fireball model also looks completely different. The compactness problem, which can be solved in the standard fireball with GRB energies of $10^{52}$–$10^{53}$ erg, does not arise for GRB energies $\sim 10^{47}$–$10^{49}$ erg. In any case, when even modest collimation of the radiation from the surface of the compact object (GRB source, XRR, or XRF) is taken into account, as is required by the angular dependence for $e-e+$ pair creation (1), this "problem" can be solved with completely different physical conditions than those proposed in [21].

2. GRB sources can, indeed, have sizes of $\sim 3 \times 10^8$ cm or less in the scenario *massive star → WR star → pre-supernova = pre-GRB*, with the total energies a factor of $10^4$–$10^6$ lower than in the standard theory, and with only a small fraction of the hard radiation that propagates within the collimation angle (2) reaching infinity. This means that old "naive" estimates of the source dimensions that follow directly from the GRB variability time scale may be quite valid if the total energies are $\lesssim 10^{49}$ erg.

Thus, the key to understanding GRBs may be that the burst energetics are much lower than in the standard fireball model.

ACKNOWLEDGMENTS

The authors thank T.N. Sokolova for editing the manuscript. This work was supported by the Russian Foundation for Basic Research (project nos. 04-02-16300, 01-02-17106, 03-02-17223), the Presidium of the Russian Academy of Sciences Basic Research Program "Non-stationary Phenomena in Astronomy," and the Ministry of Education and Science of the Russian Federation.

REFERENCES

1. S. G. Djorgovski, S. R. Kulkarni, J. S. Bloom, et al., in *Gamma-Ray Bursts in the Afterglow Era: 2nd Workshop,* Ed. by N. Masetti et al. (Springer, Berlin, 2001), p. 218; astro-ph/0107535.
2. D. A. Frail, F. Bertoldi, and G. H. Moriarty-Schiven, Astrophys. J. **565**, 829 (2002).
3. V. Sokolov, T. Fatkhullin, A. Castro-Tirado, et al., Astron. Astrophys. **372**, 438 (2001).
4. T. J. Galama, P. J. Groot, J. van Paradijs, et al., Astrophys. J. **497**, L13 (1998).
5. V. V. Sokolov, A. I. Kopylov, S. V. Zharikov, et al., Astron. Astrophys. **334**, 117 (1998); astro-ph/9802341.
6. J. S. Bloom, S. R. Kulkarni, and S. G. Djorgovski, Nature **401**, 453 (1999).
7. G. Bjornsoon, J. Hjorth, P. Jakobsson, et al., Astrophys. J. **552**, L121 (2001).
8. A. Castro-Tirado, V. V. Sokolov, J. Gorosabel, et al., Astron. Astrophys. **370**, 398 (2001).
9. D. Lazzatti, S. Covono, G. Ghisellini, et al., Astron. Astrophys. **378**, 996 (2001).
10. J. Bloom, S. Kulkarni, O. Price, et al., Astrophys. J. **572**, L45 (2002); astro-ph/0203391.
11. M. Della Valle, D. Malesani, S. Benetti, et al., Astron. Astrophys. **406**, L33 (2003).
12. J. Hiorth, J. Sollerman, P. Mшller, et al., Nature **423**, 847 (2003).
13. K. Z. Stanek, T. Matheson, P. M. Garnavich, et al., Astrophys. J. **591**, L17 (2003).
14. B. Thompsen, J. Hjorth, D. Watson, et al., Astron. Astrophys. **419**, L21 (2004); astro-ph/0403451.
15. A. Zeh, S. Klose, and D. H. Hartmann, Astrophys. J. **609**, 952 (2004); astro-ph/0311610.
16. A. Zeh, S. Klose, and D. A. Kann, astro-ph/0509299; Astrophys. J. (in press).
17. V. Sokolov, Bull. Spec. Astrophys. Obs. **51**, 38 (2001).
18. F. A. Agaronyan and L. M. Ozernoi, Astron. Tsirk., №1072 (1979).
19. L. Amati, F. Frontera, and M. Tavani, Astron. Astrophys. **390**, 81 (2002).
20. D. Q. Lamb, T. Q. Donaghy, and C. Graziani, astroph/0309463 (2003).

21. T. Piran, Nucl. Phys. (Proc. Suppl.) **70**, 431 (1999); astro-ph/9801001.
22. T. Piran, Phys. Rep. **314**, 575 (1999); astroph/9810256.
23. B. J. Carrigan and J. I. Katz, Astrophys. J. **399**, 100 (1992).
24. T. Piran, *Unsolved Problems in Astrophysics,* Ed. by J. N. Bahcall and J. P. Ostriker (Princeton Univ. Press, Princeton, 1996), p. 343; astro-ph/9507114.
25. T. Piran, Rev. Mod. Phys. **76**, 1143 (2004); astroph/0405503.
26. Y. Lithwick and R. Sari, Astrophys. J. **555**, 540 (2001).
27. G. J. Fishman and C. A. Meegan, Ann. Rev. Astron. Astrophys. **33**, 415 (1995).
28. R. D. Preece, M. S. Briggs, R. S. Mallozzi, et al., Astrophys. J., Suppl. Ser. **126**, 19 (2000).
29. E. P. Mazets and S. V. Golenetskii, *Results in Science and Technology*, Ed. by R. A. Stynyaev (VINITI, Moscow, 1987), p. 16 [in Russian].
30. D. Q. Lamb, T. Q. Donaghy, and C. Graziani, astroph/0312504 (2003).
31. D.Q. Lamb, T.Q. Donaghy, and C.Graziani, Astrophys. J. **620**, 355 (2005); astro-ph/0312634.
32. M. G. Baring and M. L. Braby, Astrophys. J. **613**, 460 (2004); astro-ph/0406025.
33. E.W. Liang, Z. G. Dai, and X. F. Wu, Astrophys. J. **606**, L25 (2004); astro-ph/0403397.
34. R. Atkins, W. Benbow, D. Berley, et al., Astrophys. J. **583**, 824 (2003); astro-ph/0207149.
35. D. Gialis and G. Pelletier, Astrophys. J. **627**, 868 (2005); astro-ph/0405547.
36. J. Heise, J. Zand, R. M. Kippen, and P. M. Woods, in *Gamma-Ray Bursts in the Afterglow Era,* Ed. by E. Costa, F. Frontera, and J. Hjorth (Springer, Berlin, 2001), p. 16; astro-ph/0111246.
37. K. A. Postnov, Usp. Fiz. Nauk **169**, 545 (1999) [Phys. Usp. **42**, 469 (1999)].
38. A. Grider, E. P. Liang, and I. A. Smith, Astrophys. J. **479**, L39 (1997).
39. R. D. Preece, M. S. Briggs, R. S. Mallozzi, et al., Astrophys. J. **506**, L23 (1998).
40. R. D. Preece, M. S. Briggs, T.W. Giblin, et al., Astrophys. J. **581**, 1248 (2002).
41. G. Ghisellini, astro-ph/0301256.
42. G. Ghirlanda, A. Celotti, and G. Ghisellini, Astron. Astrophys. **406**, 879 (2003); astro-ph/0210693.
43. F. Ryde, Astrophys. J. **614**, 827 (2004); astroph/0406674.
44. S. I. Blinnikov, A. V. Kozyreva, and I. E. Panchenko, Astron. Rep. **43**, 838 (1999).
45. M.V. Medvedev, Astrophys. J. **540**, 704 (2000); astroph/0001314.
46. T. Piran and A. Shemi, Astrophys. J. **403**, L67 (1993).
47. R. Sari, in *Fifth Huntsville Gamma-Ray Burst Symposium,* Ed. by R. M. Kippen, R. S. Mallozzi, and G. J. Fishman (American Institute of Physics, Melville, New York, 2000), AIP Conf. Proc. **526**, 504 (2000); astro-ph/002165.
48. V. Sokolov, *Gamma-Ray Burst in the Afterglow Era: 2nd Workshop,* Ed. by E. Costa et al. (Springer, Berlin, 2001), p. 132; astro-ph/0102492.
49. C. Porciani and P. Madau, Astrophys. J. **548**, 522 (2001).
50. B. Poczyrnski, *Supernovae and Gamma Ray Bursts: The Largest Explosions Since the Big Bang,* Ed. by M. Livio, N. Panagia, and K. Sahu (Cambridge Univ. Press, Cambridge, 2001), p. 1; astroph/9909048.
51. B. D. Fields, F. Daigne, M. Casse re, and E. Vangioni-Flam, Astrophys. J. **581**, 389 (2002); astroph/0107492.
52. D. Richardson, D. Branch, D. Casebeer, et al., Astron. J. **123**, 745 (2002); astro-ph/0112051.
53. Ph. Podsiadlowski, P. A. Mazzali, K. Nomoto, et al., Astrophys. J. **607**, L17 (2004); astro-ph/0403399.
54. R. Willingale, J. P. Osborne, and P. T. O'Brein, Mon. Not. R. Astron. Soc. **349**, 31 (2004); astroph/0307561.
55. E. Ramirez-Ruiz, E. Dray, P. Madau, and C. A. Tout, Mon. Not. R. Astron. Soc. **327**, 829 (2001); astroph/0012396.
56. V. G. Gorbatskii, private communication (2004).
57. L. Amati, F. Frontera, M. Vietri, et al., Science **290**, 953 (2000).
58. V. V. Sokolov, Doctoral Dissertation, http://www.sao.ru/hq/grb/team/vvs/vvs/html (2002).
59. A. Panaitescu, Astrophys. J. **556**, 1002 (2001).
60. A. Panaitescu and P. Kumar, Astrophys. J. **554**, 667 (2001); astro-ph/0010257.
61. A. Panaitescu and P. Kumar, Astrophys. J. **571**, 779 (2002); astro-ph/0109124.
62. A.M. Beloborodov, astro-ph/0405214.
63. B. Paczyrnski, Acta Astron. **42**, 145 (1992).
64. V. G. Bezchastnov, G. G. Pavlov, Yu. A. Shibanov, et al., *Gamma-Ray Bursts. 3rd Huntsville Symposium,* Ed. by C. Kouveliotou, M. F. Briggs, and G. J. Fishman, AIP Conf. Proc. (Woodbury, New York, 1996), Vol. 384, p. 907.
65. G. S. Bisnovatyi-Kogan, astro-ph/0310361.
66. G. S. Bisnovatyi-Kogan, Chinese J. Astron. Astrophys., Suppl. **3**, 489 (2003); astro-ph/0401369.
67. G. S. Bisnovatyi-Kogan, V. S. Imshennik, D. K. Nadiozhin, and V.M. Chechetkin, Astrophys. Space Sci. **35**, 23 (1975).
68. G. S. Bisnovatyi-Kogan and V. M. Chechetkin, Astrophys. Space Sci. **26**, 25 (1974).
69. G. S. Bisnovatyi-Kogan and V. M. Chechetkin, Astrophys. Space Sci. **89**, 447 (1983).
70. V. S. Berezinsky and O. F. Prilutsky, Astron. Astrophys. **175**, 309 (1987).
71. G. S. Bisnovatyi-Kogan, Astron. Zh. **48**, 813 (1970) [Sov. Astron. **15**, 941 (1970)].
72. N. V. Ardeljan, G. S. Bisnovatyi-Kogan, and S. G. Moiseenko, Astron. Astrophys. **355**, 1181 (2000).
73. N. V. Ardeljan, G. S. Bisnovatyi-Kogan, and S. G. Moiseenko, Mon. Not. R. Astron. Soc. **359**, 333 (2005).
74. S. G. Moiseenko, G. S. Bisnovatyi-Kogan, and N. V. Ardeljan, in *IAU Symp. 192: Supernovae,* Ed. by J. M. Marcaide and K. W. Weiler (2003); astro-ph/0310142.

75. V. V. Usov, Mon. Not. R. Astron. Soc. **267**, 1035 (1994).
76. C. Thompson, Mon. Not. R. Astron. Soc. **270**, 480 (1994).
77. P. Meszaros and M. J. Rees, Astrophys. J. **482**, L29 (1997).
78. R. D. Blandford, in *Lighthouses of the Universe,* Ed. byM. Gilfanov, R. A. Sunayev, and E. Churazov (Springer, Berlin, 2002), p. 381.
79. Yu. N. Gnedin andS.O.Kiikov, Mon. Not. R. Astron. Soc. **318**, 1277 (2001).
80. Yu. N. Gnedin, private communication (2004).
81. G. Calucci,Mod. Phys. Lett. A **14**, 2621 (1999).
82. S.-S. Xue, Phys. Rev. D **68**, 013004 (2003); quantph/0106076.
83. P. H. Rojas and R. E. Querts, hep-ph/0406284.
84. G. Metalidis and P. Bruno,Mon. Not. R. Astron. Soc. **341**, 385 (2003); quant-ph/0207153.
85. K. Hurley, T. Cline, and E. Mazets, Nature **397**, 41 (1999).
86. E. P. Mazets, T. L. Cline, and R. L. Aptekar, astroph/0502541 (2005).
87. V. V. Usov, Astrophys. J. **550**, L179 (2001).
88. V. V. Usov, Astrophys. J. **559**, L135 (2001).
89. J. P. Norris; astro-ph/0307279 (2003).
90. J. P. Norris and J. T. Bonnel, astro-ph/0312279 (2003).
91. K. A. Postnov, Usp. Fiz. Nauk **169**, 545 (1999) [Phys. Usp. **42**, 469 (1999)].
92. L. Piro, E. Costa,M. Feroci, et al., Astron. Astrophys. Space Sci. **138**, 431 (1999).
93. A. Yoshida, M. Namiki,D. Yonetoku, et al., Astrophys. J. **557**, 27 (2001).
94. L. Piro, G. Garmire, M. Garcia, et al., Science **290**, 955 (2000).
95. A. Antonelli, L. Piro, M. Vietri, et al., Astrophys. J. **545**, L39 (2000).
96. G. Ghisellini, D. Lazzati, E. Rossi, and M. J. Rees, Astron. Astrophys. **389**, 33 (2002); astroph/0205227.
97. G. B. Taylor, D. A. Frail, E. Berger, and R. Kulkarni, Astrophys. J. **609**, L1 (2004); astro-ph/0405300.

*Translated by D. Gabuzda*